\documentclass[a4paper, 10pt, conference]{ieeeconf}      

\IEEEoverridecommandlockouts                              

\usepackage{listings}
\usepackage{hyperref}
\usepackage{fontawesome}
\newtheorem{defi}{Definition}
\def\accd{\mathsf{nsb}}
\def\acce{\mathsf{nsb_e}}

\def\ufp{\mathsf{ufp}}
\def\ulp{\mathsf{ulp}}
\def\ufpe{\mathsf{ufp_e}}
\def\ulpe{\mathsf{ulp_e}}

\usepackage{graphics} 
\usepackage{epsfig} 
\usepackage{mathptmx} 
\usepackage{times} 
\usepackage{amsmath} 
\usepackage{amssymb}  

\title{\LARGE \bf
Constrained Precision Tuning
}
\author{Dorra Ben Khalifa$^{1}$ and Matthieu Martel$^{1, 2}$
\thanks{$^{1}$ The authors are with the LAMPS Laboratory of the university of Perpignan, 52 Avenue Paul Alduy, Perpignan, 66100, France
 {\tt\small dorra.ben-khalifa@univ-perp.fr} {\tt\small matthieu.martel@univ-perp.fr}}%
\thanks{$^{2}$ Matthieu Martel is also  with the Numalis company, 265 Avenue des \'Etats du Languedoc, Montpellier, 34000, France.}%
}
\begin{document}
\maketitle
\thispagestyle{empty}
\pagestyle{empty}

\begin{abstract}
Precision tuning or customized precision number representations is emerging, in these recent years, as one of the most promising techniques that has a positive impact on the footprint of programs concerning energy consumption, bandwidth usage
and computation time of numerical programs. In contrast to the uniform precision, mixed precision tuning assigns different finite-precision types to each variable and arithmetic operation of a program and offers many additional optimization opportunities. However, this technique introduces new challenge related to the cost of operations or type conversions which can overload the program execution after tuning. In this article, we extend our tool POP (Precision  OPtimizer), with efficient ways to limit the number of drawbacks of mixed precision and to  achieve best compromise between performance and memory consumption. On
a popular set of tests from the FPBench suite, we discuss the results obtained by POP.
\end{abstract}
     
\section{Introduction}\label{sec1}
In recent years, precision tuning to improve the performance metrics is emerging as a new trend to save the resources on the available processors, especially when new error-tolerant applications are considered~\cite{CA20}. By way of illustration,  many applications can tolerate some loss in quality during computation, as in the case of media processing (audio, video and image), data mining, machine learning, etc. In addition, as almost all numerical computations are performed using floating-point operations to represent real numbers~\cite{IEEE754}, the precision of the related data types should be adapted in order to guarantee the desired overall rounding error and to strengthen the performance of programs. For instance, using FP32 single precision formats is often at least twice as fast as the FP64 double precision ones on most modern processors~\cite{BaboulinBDKLLLT09}. Consequently, the natural question that arises is how to obtain the best precision/performance trade-off by allocating some program variables in low precision (e.g. FP16 and FP32) and by using high precision (e.g. FP64 and FP128) selectively. This process is also called, mixed-precision tuning.

Let us precise that precision tuning is not a simple task limited to changing the data type in the source code with the Find-and-Replace button of any text editor. It is a more complex technique which analyzes the semantics of the programs and presents several challenges both architectural and algorithmic. For this reason, various tools~~\cite{ChiangBBSGR17,DarulovaHS18,GR18,KSWLB19,LHSL13,RGNNDKSBIH13} have been proposed to help developers select the most appropriate data representations. Such tools may integrate different approaches but their common goal is still to automatically or semi-automatically adapt an original code given in higher precision to the selected lower precision type. However, the common point to all the existing techniques
 is that they follow a trial-and-error strategy: they change the data types of some variables
 of the program and evaluate the accuracy of the result and depending on what is obtained they change more or less data types and repeat the process. At this level, we present an important difference relating the terms precision and accuracy
 that are often confused, even though they have significantly different meanings. Here, we
 call precision a property of a number format that refers to the amount of information used to represent a number. Better or higher precision means more numbers can be represented,
 and also means a better resolution. Otherwise, the term accuracy denotes how close a
 floating-point computation comes to the real value \cite{Mar17}: a bound on the absolute error $|x - \widehat{x}|$ between the represented $\widehat{x}$ value and the exact value $x$ that we would have in the exact arithmetic.

The POP tool~\cite{BKM21,KMA21} proposes a novel static
technique based on a semantic modelling of the propagation of the numerical errors
throughout the code formulated with two methods. The first method consists of generating an Integer
Linear Problem (ILP) from the program. Basically, this is done by reasoning on the most significant bit and
the number of significant bits of the values which are integer quantities. The integer solution to this problem,
computed in polynomial-time by a classical linear programming solver, gives the optimal data types at bit-level. The second method proposes a finer set of semantic equations which does not reduce directly to an ILP problem. So, we
use the policy iteration (PI) technique to find a solution. Let us note that the originality of both methods is to find directly the minimal number of bits needed, known as bit-level precision tuning, at each control point of the variables of the program.

In this article, we focus on improving the efficiency of POP by experimenting several optimization criteria to our system of constraints 
in order to achieve the best compromise between performance and memory saving. The first criteria is related to minimize the number of formats 
in the tuned programs.  The second criteria correponds to minimize the number of bits needed for each operation performed in programs. 
The third optimization function aims to avoid  type conversions of the same variables in the same program.

The remainder of this article is as follows. Section~\ref{sec2} describes the static approach inside the POP tool. Section~\ref{sec3} point out the new optimization criteria related to the type conversions of tuning. Experimental results are presented in Section~\ref{sec4}. Section~\ref{sec5} discusses the state-of-the-art tools for precision tuning before concluding in Section~\ref{sec6}.

\section{POP: Precision OPtimizer}\label{sec2}

In this section, we start by describing the necessary  background to understand the technique behind our tool POP (Section~\ref{back}). Next, we highlight the main architecture of POP in Section~\ref{POP}. Section~\ref{example} presents a motivating example of tuning with POP.
\subsection{Background}\label{back}
\begin{figure}[t]
  \begin{center}
   \includegraphics[width=7cm]{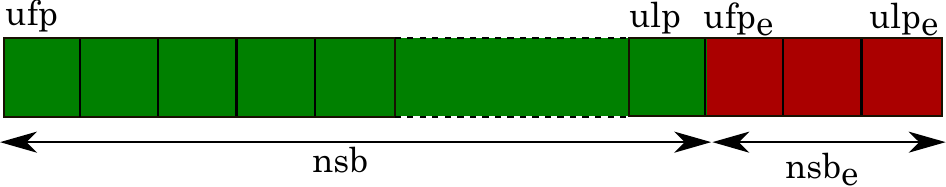} 
 \caption{Schematic representation of $\ufp$, $\accd$ and $\ulp$ for values and errors.}\label{numberX}
 \end{center}		
 \end{figure}

POP manipulates numbers for which we know their unit in the first place denoted by $\ufp$ (Definition~\ref{defiUfp}), their number of significant bits, denoted by $\accd$ (Definition~\ref{defiNsb}) and their unit in the last place denoted by $\ulp$ (Definition~\ref{defiUlp}). We also assume that the constants occurring in the source codes are exact and we bound the errors introduced by the finite precision computations. These functions are defined hereafter and more intuitive presentation is given in Figure~\ref{numberX}.

 \begin{defi}[Unit in the First Place]\label{defiUfp}\normalfont
    The unit in the first place of a real number $x$, denoted by $\ufp(x)$, and possibly encoded up to some rounding mode by a floating-point or a fixed-point number is given in Equation~(\ref{ufp}). This function is independent of the representation of $x$:
      \begin{equation}\small
         \label{ufp}
         \ufp(x) =\left\{ \begin{array}{lr}
            \min \{i \in \mathbb{Z} : 2^{i+1} > \lvert x \rvert\} = \lfloor  \log_2( \lvert x \rvert ) \rfloor& \text{if} \  x \neq 0, \\
                      0 & \text{if} \  x = 0.
             \end{array}\right.
         \end{equation}
        \end{defi}

 \begin{defi}[Number of Significant Bits]\label{defiNsb}\normalfont
    Intuitively, $\accd(x)$ is the number of significant bits of $x$.
    Let $\hat{x}$  the approximation of $x$ in finite precision and let $\varepsilon(x) = \lvert x - \hat{x} \rvert$ be the absolute error. Following Parker~\cite{parker},  if  $\accd(x) = k$, for $x\not=0$, then
         \begin{equation}\label{bound}
  \varepsilon(x) \le 2^{\ufp(x) - k + 1} \enspace .
  \end{equation}
  \end{defi}
  
   \begin{defi}[Unit in the Last Place]\label{defiUlp}\normalfont
  The unit in the last place of a number $x$ denoted by $\ulp(x)$ is defined below in Equation~(\ref{ulp}). It depends on the unit in the first place $\ufp(x)$ and the number of significant bits $\accd(x)$: 
      \begin{equation}\label{ulp}
      \ulp(x) = \ufp(x)- \accd(x)+1 \enspace.
      \end{equation}
  \end{defi}

 For example, if  $x= 2.75$ then, following equations~\ref{ufp} to~\ref{ulp}, we have $\ufp(x)= 1$, $\accd(x) = 4$ and consequently $\ulp(x) = -2$. Note that we define in the same manner $\ufpe(x)$ and $\acce(x)$ respectively the $\ufp$ and $\accd$ of the error on a number $x$ (see Figure~\ref{numberX}) which are used to describe the error propagation through the computations and to compute either a carry bit can be propagated through a computation or not. In practice, POP implements an optimized carry bit function, denoted by $\xi$, that adds an extra bit in some arithmetic operation only if the  errors of the operands can overlap. Definition~\ref{newXi} highlights this function.

 \begin{defi}[Carry Bit Function]\label{newXi}\normalfont Let $x$ and $y$ two operands of some operation which result is $z$. The optimized $\xi$ function is given as shown in Equation~(\ref{xi}): if the $\ulpe$ of one of the two operands error $\varepsilon(x)$ or $\varepsilon(y)$ is greater than the $\ufpe$ of the other one (or conversely) then the numbers $x$ and $y$ are not aligned and consequently $\xi = 0$ (otherwise $\xi = 1$). The optimized $\xi$ function is given by
    \begin{equation}\label{xi}\small
        \xi(z)(x, y) = \begin{cases} 0 &\ulpe(x)  \ge \ufpe(y), \\
    0 &\ulpe(y)  \ge \ufpe(x), \\
            1 & \text{otherwise.} \end{cases}
    \end{equation}
    \end{defi}
Let us state that the carry bit function has an effect on the linearity of the constraints that we will highlight in the next section.
\subsection{POP Outline}\label{POP}
 From a grammar, POP parses the input source codes using  the ANTLR framework~\cite{antlr}. Next, it performs a range determination consisting in launching the execution of the program a certain number of times in order to determine dynamically the range of variables. More precisely, this is done by reasoning on the weight of the most significant bit already defined in Equation~(\ref{ufp}).

   The new features of POP, in contrast to its former introduction in~\cite{KMA19,KM19,KM20,SMT21}, is to solve an ILP problem generated from the program source code. Next, the optimal solution computed by a
   classical LP solver (we use GLPK\footnote{https://www.gnu.org/software/glpk/} in practice) gives the optimized data types that satisfy the user accuracy requirement in a polynomial-time and with respect to some optimization function.  We remind the reader that the goal of this article is to introduce several optimization functions and to  evaluate the performance of tuning in terms of mixed-precision and quality of analysis. However, we must precise that we have over-approximated the carry bit propagation throughout the computations in the ILP formulation because of the $\min$ and $\max$ operators that arise when solving Equation~(\ref{xi}).
    For this purpose,  POP implements an optimization of the previous ILP method by introducing a second set of semantic equations. These new equations make it possible to tune even more the precision by being less pessimistic on the propagation of carries in
   arithmetic operations. So, we use the policy iteration (PI) method~\cite{CGGMP05} to find efficiently a solution. We refer the reader to~\cite{KMA21} for a detailed explanation about our method. 
   
   Concerning the complexity of the analysis performed by POP, in practice, the number
   of variables and constraints is linear in the size of the program and consequently the complexity to analyze a program of size $n$ is equivalent to that of solving a system of $n$ constraints in our language of constraints. In addition, POP is able to analyze large programs ($\approx$ KLOC) but the only limitation is related to the size of the problem accepted by the solver.

\subsection{Motivating Example}\label{example}
\begin{figure*}[t]\small
  \noindent\hrule
   \begin{center}
   ~\hspace{-1cm}
   \begin{minipage}[c]{0.39\linewidth}
 \begin{lstlisting}[mathescape=true]
dt$^{\ell_2}$ = 0.5$^{\ell_0}$; invdt$^{\ell_5}$ = 0.5$^{\ell_3}$;
kp$^{\ell_8}$ = 9.4514$^{\ell_6}$; ki$^{\ell_{11}}$ = 0.69006$^{\ell_{9}}$;
kd$^{\ell_{14}}$ =2.8454$^{\ell_{12}}$; c$^{\ell_{17}}$ = 5.0$^{\ell_{15}}$;
m$^{\ell_{20}}$ = 8.0$^{\ell_{18}}$; e$^{\ell_{23}}$ = 0.0$^{\ell_{21}}$; 
p$^{\ell_{26}}$ = 0.0$^{\ell_{24}}$; i$^{\ell_{29}}$ = 0.0$^{\ell_{27}}$;
d$^{\ell_{32}}$ = 0.0$^{\ell_{30}}$; r$^{\ell_{35}}$ = 0.0$^{\ell_{33}}$;
m1$^{\ell_{39}}$ = m$^{\ell_{37}}$; eold$^{\ell_{42}}$= 0.0$^{\ell_{40}}$; 
t$^{\ell_{45}}$ = 0.0$^{\ell_{43}}$;
while (t<100.0) {
  e$^{\ell_{56}}$ = c$^{\ell_{51}}$-$^{\ell_{54}}$ m1$^{\ell_{53}}$;
  p$^{\ell_{63}}$ = kp$^{\ell_{58}}$ *$^{\ell_{61}}$ e$^{\ell_{60}}$;
  i$^{\ell_{76}}$ = i$^{\ell_{65}}$ +$^{\ell_{74}}$ ki$^{\ell_{67}}$ *$^{\ell_{70}}$ dt$^{\ell_{69}}$ *$^{\ell_{73}}$ e$^{\ell_{72}}$;
  d$^{\ell_{89}}$ = kd$^{\ell_{78}}$ *$^{\ell_{81}}$ invdt$^{\ell_{80}}$ *$^{\ell_{87}}$ e$^{\ell_{83}}$-$^{\ell_{86}}$ eold$^{\ell_{85}}$;
  r$^{\ell_{99}}$ = p$^{\ell_{91}}$ +$^{\ell_{94}}$ i$^{\ell_{93}}$ +$^{\ell_{97}}$ d$^{\ell_{96}}$;
  m1$^{\ell_{108}}$ = m1$^{\ell_{101}}$ +$^{\ell_{106}}$ 0.01$^{\ell_{102}}$ *$^{\ell_{105}}$ r$^{\ell_{104}}$;
  eold$^{\ell_{112}}$ = e$^{\ell_{110}}$;
  t$^{\ell_{119}}$ = t$^{\ell_{114}}$ +$^{\ell_{117}}$ dt$^{\ell_{116}}$;
}$^{\ell_{120}}$ ;
require_nsb(m1,12)$^{\ell_{122}}$;   
 \end{lstlisting}
   \end{minipage}
    \qquad 
   \begin{minipage}[r]{0.45\linewidth}\small
 \begin{lstlisting}
  dt|12|} = 0.5|12|; invdt|12| = 0.5|12|;
  kp|12| = 9.4514|12|;  ki|12| =0.69006|12|;
  kd|12| = 2.8454|12|; c|14| = 5.0|14|;
  m|15| = 8.0|15|; e|0| = 0.0|0|;
  p|0| = 0.0|0|; i|13| = 0.0|13|;
  d|0| = 0.0|0|; r|0| = 0.0|0|;
  m1|15| = m|15|; eold|12| = 0.0|12|;
  t|13| = 0.0|13|;
  while (t<100.0) {
    e|12| = c|14| -|12| m1|15|;
    p|13| = kp|12| *|13| e|12|;
    i|12| = i|13| +|12| ki|9| *|10| dt|9| *|11| e|10|;
    d|12| = kd|10| *|11| invdt|10| *|12| e|12| -|11| eold|12|;
    r|12| = p|13| +|12| i|11| +|12| d|10|;
    m1|12| = m1|14| +|12| 0.01|8| *|9| r|8|;
    eold|12| = e|12|;
    t|12| = t|13| +|12| dt|6|;
  } ;
  require_nsb(m1,12);
 \end{lstlisting}
   \end{minipage}
 \end{center}
 \noindent\hrule
 \vspace{0.1cm}
 \caption{\label{listing2} Top left: PID source program annotated with labels. Top right: PID program tuned with POP.
 }
\end{figure*}
To better explain the ILP nature of the precision tuning problem in POP, we consider the PID controller program~\cite{DMC15} of Figure~\ref{listing2}. This algorithm is widely used  in embedded and critical systems e.g. aeronautic and avionic systems. The main feature of this program is to keep a physical parameter $m$ at a specific value known as the setpoint . In other
words, it tries to correct a measure by maintaining it at a defined value. The original
PID program is depicted in the left hand side of Figure~\ref{listing2} whereas the optmized program by POP is given in the right hand side corner. 

Some points can be highlighted about this example. For instance, the variables ${dt}$ and ${invdt}$ are  both initialized respectively to the value $0.5$, annotated with their control points thanks to the following annotations ${dt^{\ell_{2}} = 0.5^{\ell_{0}}}$ and ${invdt^{\ell_{5}} = 0.5^{\ell_{3}}}$ in the left hand side of Figure~\ref{listing2}. Also, the statement $require\_nsb(m1,12)^{\ell_{122}}$ at Line $19$ informs POP that the user wants to get $12$ significant bits on variable $m1$ and so, $\accd(m1) = 12$. Next, POP reduces the problem to an ILP formulation by generating a set of constraints on all the labels of programs. Finally, the optimal solution computed by a the GLPK solver~\cite{GLPK}  gives the optimized data types that satisfy the user accuracy requirement in a polynomial-time as shown in the right hand side of Figure~\ref{listing2}. By way of illustration, the results obtained on Line $10$ of the PID program says that for $\accd(m1) = 12$  bits,  the number of significant bits needed for variable $e$ is $12$ bits, the variable $c$ is computed with $14$ bits whereas the operator $-$ is computed with $12$ bits, etc.

\section{Guiding Tuning with Constraints}\label{sec3}
In this section, we introduce the different cost functions used by POP to obtain the best compromise between performance and memory saving. 
Recall that POP generates a set of constraints whose solution gives the tuning (See Section~\ref{POP}.) Our new optimization criteria are then 
expressed as cost functions that the solver has to optimize. Below we propose three optimization criteria related to the largest data type,  
the number of bits needed for each operation and the prohibition of type conversions. 

Let us remark that one may combine the cost functions introduced in this article and try and optimze the accuracies at any
control point globally. While encompassing the cost functions presented in previous work~\cite{KMA21} in terms of tuning, this method suffers from several drawbacks. First, it increases the number of constraints and variables of the system. Second, it over-constraints the system which makes the solver fail more
often and  finally it slows down the tuning time.

Let us also remark that, compared to other approaches such as the ones based on delta-debugging (see Section \ref{sec6}),
POP approach which is based on solving a system of constraints allows one to define easily many optimization criteria
such as the ones introduced below without modifying significantly the tool and without incereasing the combinatory of the problem.

\subsection{Minimize the Number of the Largest Data Type}
\label{seclargedt}

The purpose of this cost function is to answer the following question: \textit{What is the minimal 
number of bits of the greatest format needed in the program in order to ensure some accuracy on the result?}
For example, by this technique one may answer to questions such as \textit{is it possible to obtain a result with $18$ significant bits using only, e.g. single precision numbers (FP32)? }

More formally, let $Lab$ denotes the set of labels of the program and
let $T: Lab \rightarrow \mathbb{N}$ be a tuning assigning to each control point $\ell \in Lab$ an integer precision. A tuning is correct if it satisfies the system of constraints generated by POP~\cite{KMA21} including the accuracy requirement on the result and we denote $\mathcal{T}$ the set of correct tunings. The cost function for maximal precision $MP$ that we aim to compute is given as shown in Equation~(\ref{cost1}).
\begin{equation}\label{cost1}
    MP = \min_{T \in\mathcal{T}} \left\{\max_{\ell \in Lab}T(\ell) 
    \right\}  \enspace .
\end{equation}
Minimizing the largest format may enable one to use a processor with limited formats (e.g. only single precision). 
In former work, our cost function was  to minimize the sum of the precision of the assigned variables in the program, i.e. $\sum_{\ell \in Lab}T(\ell)$.
However, this may lead to cases where some variables have large formats and others small ones (e.g. from FP16 half precision to FP64 double)  which makes difficult hardware optimizations \cite{KMA19}.

\subsection{Minimize of the Operations Number of Bits}
\label{secop}

Our second cost function $MOp$ focuses on the operators instead of the variables of the program. We aim at
minimizing only the number of bits used in the arithmetic operations, without considering what is used for variables.
The interest is to minimize the hardware needed to run the programs. Also, this optimization is particularly
relevant for circuit implementations, e.g. using FPGAs~\cite{GC15}. 

Formally, let $Op\subseteq Lab$ be the subset of labels attached to operators (additions, multiplications, elementary functions, etc.) Here, we aim at computing
\begin{equation}\label{cost2}
    MOp = \min_{T \in\mathcal{T}} \left\{\sum_{\ell \in Op}T(\ell) 
    \right\} \enspace .
\end{equation}

In the present work, we assign the same weight to each operation (i.e. its number of bits). However, it would be
interesting to assign different weights, for instance to take into account that a multiplication is more costly than an 
addition at the hardware level (same for elementary functions.)

\subsection{Avoid Type Conversions}
\label{seccf3}

Mixed precision tuning, as done by POP, offers the advantage of optimizing the precision of a variable at each of its 
occurences. However, from a performance point of view, this introduces type conversions  which may slow down the programs.
Let $\mathcal{V}:\ Var\rightarrow \wp(Lab)$ be a function mapping each variable $x$ of a program to the set of labels
corresponding to the occurrences of $x$ and let $Dom(\mathcal{V})$ denote the definition domain of $\mathcal{V}$.  
We add a mode in POP which enforces it to produce an uniform tuning by adding the constraints
\begin{equation}\label{cost3}
  \forall x\in Dom(\mathcal{V}),\ \forall \ell_1,\ \ell_2\in\mathcal{V}(x),\ T(\ell_1)=T(\ell_2) 
    \enspace .
\end{equation}
Let us remark that, in this mode, POP still achieve bit-level precision tuning. However this tuning is uniform and only one precision
is returned for each variable which avoids type conversions. As previously mentionned, let us also underline the fact that since
POP is based on a system of constraints, assigning to it new optimization objectives can be done easily, without a deep refactoring
of the tool.

\normalfont
\section{Experimental Evaluation}\label{sec4}
\normalsize
We evaluate the performance of POP  on a standard benchmark set for floating-point analysis (Section~\ref{secbench}), and compare the results of precision tuning using the different cost functions already defined in Section~\ref{sec3}. Note that POP and all the data and results presented in this article (Section~\ref{eval}), are publicly available under GPL-3.0 License\footnote{ \faGithub  \url{https://github.com/benkhelifadorra/POP-v2.0}}. Also, the experiments are ran on an Intel Core i5-8350U 1.7GHz Linux machine with 8 GB RAM.

 \begin{table}[t] \scriptsize
 \centerline{
 \begin{tabular}{lcccccccc}
 \hline
 &&&&&&&&\\
 Prog. & \multicolumn{4}{c}{Require 8 bits} & \multicolumn{4}{c}{Require 12 bits}\\
 &&&&&&&&\\
         & MP & FP16 & FP32 & FP64 & MP & FP16 & FP32 & FP64 \\
 Acc. & \textbf{9} & 18 & 0 & 0  & \textbf{13} & 3 & 15 & 0\\
 Odo. & \textbf{13} & 22 & 7 & 0  & \textbf{17} & 0 & 29 & 0\\
 Pend. & \textbf{17} & 12 & 1 & 0  & \textbf{21} & 0 & 13 & 0\\
 PID  & \textbf{12} & 17 & 2 & 0  & \textbf{16} & 0 & 19 & 0\\
 RK & \textbf{10} & 22 & 0 & 0  & \textbf{14} & 0 & 22 & 0\\
 Trap. & \textbf{14} & 5 & 10 & 0  & \textbf{18} & 0 & 15 & 0\\
 &&&&&&&\\ 
 & \multicolumn{4}{c}{Require 16 bits} & \multicolumn{4}{c}{Require  32 bits}\\
 &&&&&&&\\ 
 & MP & FP16 & FP32 & FP 64 & MP & FP16 & FP32 & FP 64 \\
 Acc. & \textbf{17} & 0 & 18 & 0  & \textbf{25} & 0 & 15 & 3\\
  Odo. & \textbf{21} & 0 & 29 & 0  & \textbf{37} & 0 & 0 & 29\\
  Pend. & \textbf{25} & 0 & 12 & 1  & \textbf{33} & 0 & 10 & 3\\
  PID  & \textbf{20} & 0 & 19 & 0  & \textbf{28} & 0 & 10 & 9\\
  RK & \textbf{18} & 0 & 22 & 0  & \textbf{26} & 0 & 18 & 4\\
  Trap. & \textbf{22} & 0 & 15 & 0 & \textbf{30} & 0 & 3 & 12\\
  &&&&&&&\\
  \hline\\
 \end{tabular}}
 \caption{\label{tabcf1}Minimization of the worst precision in function of the accuracy requirement. 
 Maximal precision (MP) computed by POP followed by the number of FP16, FP32 and FP64 variables.}
 \end{table}
 
 \normalsize

 \begin{figure}[b]
     \centerline{\includegraphics[width=1.05\columnwidth]{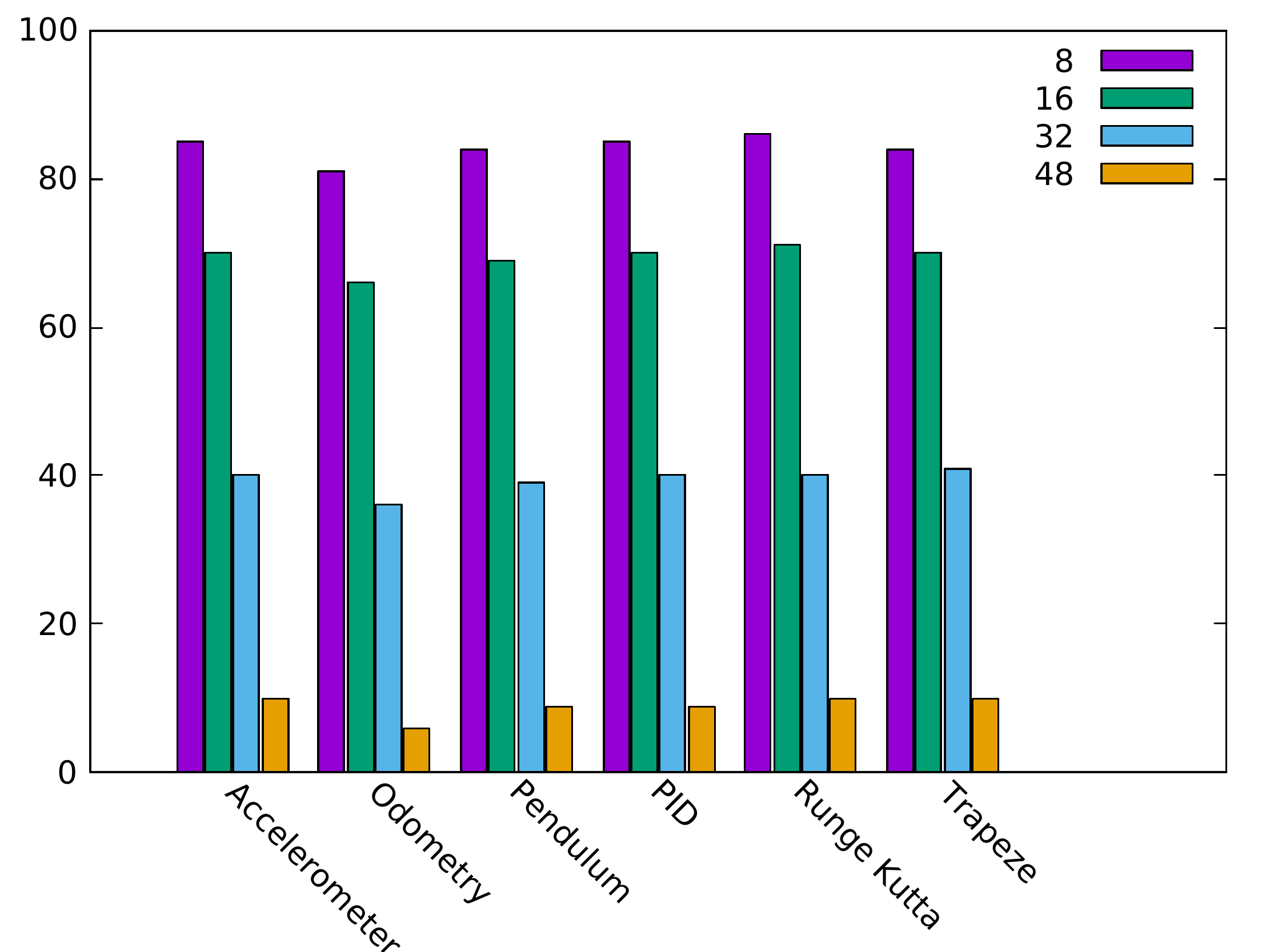}}
     \caption{Percentage of reduction after tuning by POP of the total number of bits for the operators occurring in our test programs
     in function of the accuracy requirement on the result.}
      \label{figop}
    \end{figure}

    \begin{table}[tb]  \scriptsize
        \centerline{
    \begin{tabular}{cccccccccccc}
        \hline
    &&&&&&&&\\
Prog.&        \multicolumn{2}{c}{8 bits} & \multicolumn{2}{c}{16 bits} & \multicolumn{2}{c}{24 bits} & \multicolumn{2}{c}{32 bits} \\
    &&&&&&&&\\
Acc.&    85 \% & 85 \% & 71 \% & 71 \% & 57 \% & 57 \% & 42 \% & 42\%\\
Odo. &   86 \% & 81 \% & 76 \% & 73 \% & 65 \% & 62 \% & 54 \% & 51\%\\
Pend.&    85 \% & 84 \% & 72 \% & 71 \% & 59 \% & 57 \% & 46 \% & 44\%\\
PID&    86 \% & 85 \% & 73 \% & 73 \% & 61 \% & 60 \% & 48 \% & 48\%\\
RK&    85 \% & 85 \% & 71 \% & 71 \% & 58 \% & 57 \% & 44 \% & 44\%\\
Trap.&    80 \% & 78 \% & 67 \% & 65 \% & 53 \% & 52 \% & 40 \% & 38\%\\
    &&&&&&&&\\
\hline\\
\end{tabular}}
\caption{\label{tabcost3}Memory savings (in percentage) on the number of bits needed to store variables, in mixed or uniform precision, in function of the accuracy requirement, for our test programs.}
\end{table}
\normalfont

\subsection{Benchmarks}
\label{secbench}
\normalfont
Our cost functions are experimented on several FPBench\footnote{\url{https://fpbench.org/}} benchmarks. FPBench develops standards for describing floating-point benchmarks and for measuring their accuracy~\cite{fpbench}. We have selected programs from the embedded systems, IoT and numerical analysis fields. 
\begin{itemize}
\item \textbf{Accelerometer}~\cite{KM19}: this program comes from the IoT field and  measures the angle of inclination of an object.
\item \textbf{Odometry}~\cite{DMC17}: an example taken from robotics which concerns the computation of the position($x$,$y$) of a two wheeled robot by odometry.
\item \textbf{Pendulum}~\cite{KMA21}: models the movement of a simple pendulum without damping.
\item \textbf{PID Controller}~\cite{DMC15}: is a widely used algorithm in embedded and critical systems e.g. aeronautic and avionic systems. This program was highlithed in Section~\ref{example}.
\item \textbf{Runge Kutta method}~\cite{atkinson1991}:  is an effective and widely used method for solving the initial-value problems of differential equations.
\item{\textbf{The trapezoidal rule}}~\cite{atkinson1991}: a well known algorithm in numerical analysis which approximates the definite integral $\displaystyle \int_{a}^{b} f(x) \, \mathrm{d}x$.
\end{itemize}

\subsection{Evaluation}\label{eval}

In this section, we present the results obtained with POP on the sample programs described in Section~\ref{secbench} for
the cost functions introduced in Section~\ref{sec3}.

Table~\ref{tabcf1} presents the results of the cost function  concerning the minimization of the largest data type already presented in Section~\ref{seclargedt}. In this experiment, we consider four accuracy requirements for the results of our test
programs: $8$, $12$, $16$ and $32$ significant bits. For each program and requirement, we display the maximal precision $MP$
as defined in Equation~(\ref{cost1}), as well as the number of variable in half, single and double precision (FP16, FP32 and FP64
respectively). For example, for the accelerometer program with 
an accuracy requirement of $12$ bits, we find that $MP=13$, meaning that $3$ variables may be set in half precision and that
the remaining $15$ variables may be set in single precision. A first observation is that the $MP$s are not intuitive and should be difficult to obtain without POP, either by hand or by tuning tools based on delta-debugging (because of the theoretical complexity which makes the bit-level precision tuning untracktable for this
class of tools, see related work in Section~\ref{sec5}). A second observation is that POP often finds $MP$s close to the
accuracy requirement of the result, which means that the optimization is important. Not surprisingly, the $MP$ increases as
the required number of significant bits on the result increases.

Our second experiment concerns the cost function of Section \ref{secop} related to the size of the arithmetic operators.
Our results are summarized in Figure~\ref{figop}. The histogram gives percentages of optimization for the total
number of bits needed for the operations. The percentage is computed with respect to an initial number of bits
corresponding to all the operations done in FP64 double precision (which corresponds to 100\%.) The first four
bars are for the accelerometer program, the next four bars are for the odometry program, etc.
The four bars dedicated to a program correspond to the gains obtained for the accuracy requirements set on the results: $8$, $16$,
$32$ and $48$ bits. For example, for the accelerometer program, the total number of bits for the operations is reduced by  
85\% and 70\% for requirements of $8$ and $16$ bits respectively.

A first observation is that POP is able to reduce very significantly the number of bits needed when small $\accd$ are required on
the outputs (around 80\% and 70\% of reduction for $\accd = 8$ or $\accd = 16$  bits). Also,  we observe that when large $\accd$ quantities are required (e.g. $48$ bits is close to the $53$ bits of the double precision), POP is able to reduce the size of the operators by around 10\%.

Finally, our last experiment, summed up in Table \ref{tabcost3} is for uniform precision as defined in Section \ref{seccf3}.
We consider four accuracy requirements for the results of our test programs: 8, 16, 24 and 32 bits. For each of these requirements,
we run POP in mixed-precision mode and then in uniform precision mode. We display the percentages of optimization on the total number
of bits used by the variables. Again, 100\% corresponds to all the variables in double precision. In Table \ref{tabcost3},
two consecutive values are for the same requirement, in mixed and uniform precision respectively. For example, for the odometry program
and for a requirement of 8 bits, the savings in number of bits obtained by POP are of 86\% in mixed precision and of
81\% in uniform precision. Similarly, for the same program and for a requirement of 32 bits, the saving are of
54\% and 51\% respectively. From Table \ref{tabcost3}, we may observe that POP still optimizes well the programs in uniform precision mode. This is mainly due to the fact that our sample programs do not use the same variables many times (see, for example
the code of the PID program in Figure \ref{listing2}.) A second remark is that bit-level precision tuning makes it
possible to obtain important memory savings, even in uniform mode.

\section{Related Work}\label{sec5}
The last few years have seen a wealth of precision tuning tools. In this section, we discuss the strengths and shortcomings of each tool. Besides, we classify these approaches into
two categories: static methods that extract additional knowledge from the program source
code without executing it with input data and dynamic methods that involve the profiling of the target application to extract
pieces of information by running the original version of the program.
\subsection{Static Analysis Tools}\label{sec51}

Rosa~\cite{DarulovaK17} is a source-to-source compiler which takes as input a real-valued program
with error specifications and synthesizes code over an appropriate floating-point (FP32,
FP64, FP128, and an extended format with 256 bit width) or fixed-point data type (8, 16, 32
bit) which fulfills the specification. Unlike POP which is able to ensure mixed precision tuning on programs containing
expressions, loops, conditionals and even arrays,
Rosa handles conditional statements soundly and  assigns only uniform preicision to
the variables of their programs. In addition, FPTuner~\cite{ChiangBBSGR17} exposes a user-defined threshold for the
amount of type casts that the tool may insert into the code. Let us state that the approach
deployed by FPTuner is close the static technique of POP, especially in the
constraint generation step. However, it relies on a local optimization procedure by solving
quadratic problems for a given set of candidate data types. Contrarily to POP , FPTuner
is limited to straight-line programs.  Moreover, the TAFFO tool~\cite{CherubinCCBA20} is a LLVM-based tool-chain. Its strategy is to collect statically annotations from the source code and it converts them into LLVM-IR metadata
with the goal to replace floating-point operations with fixed-point operations to the extent
possible. This analysis is used to project on the output the error introduced by each fixed-point instruction.  In contrast to TAFFO, POP is able to return solutions at bit-level suitable for the IEEE754 floating-point arithmetic, the
fixed-point arithmetic and the MPFR library for non-standard precision. Nevertheless, the static tool Daisy~\cite{DarulovaHS18} is able to provide a mixed precision solution that considers both
floating-point and fixed-point data making it generally applicable to both scientific com-
puting and embedded applications. However, Daisy does not address conditional-based
programs. 

\subsection{Dynamic Analysis Tools}\label{sec52}
Precimonious~\cite{RGNNDKSBIH13} is a dynamic automated search-based tool that leverages the LLVM framework to tweak variable declarations to build and prototype mixed-precision configurations within a given error threshold. It  is based on the delta-debugging algorithm search~\cite{Zeller} which guarantees to find a local $1$-minimum if one exists. A configuration is said to be $1$-minimal if lowering any additional variable (or function call) leads to a configuration that produces an inaccurate result, or is not faster than the original program. Unlike POP which optimizes all the variables of
the program, Precimonious optimizes only the precision of declared variables. It uses
external description files (JSON or XML) to declare which variables in the source code
should be explored and which data types have to be investigated. Moreover, it estimates
round-off errors by dynamically evaluating the program on several random inputs. For this reason,  
the Blame Analysis technique~\cite{Rubio-Gonzalez016} aims  at reducing the space of variables of Precimonious. It performs  shadow execution to identify variables that are numerically insensitive and which can consequently be excluded from the search space before tuning. The analysis finds a set of variables that can be in single precision, while the rest of the variables are in double precision. However, the output configurations may or may not improve performance, so to use the analysis in practice one must perform runs of the program to determine which configurations actually improve performance. Another dynamic tool sharing some methodologies of Precimonious is called PROMISE~\cite{GraillatJPFL19}. It modifies automatically the precision of variables taking into account an accuracy requirement on the computed result. Based on the delta-debugging search algorithm which reduces the search space of the possible variables to be converted, it provides a subset of the program variables which can be converted from FP64 to FP32 only. Meanwhile, PROMISE is able to tune programs only in FP$32$ single precision and it remains a time-intensive tool. HiFPTuner~\cite{GuoR18}  is another extension of Precimonious which uses a hierarchical search approach. It combines a static analysis to create the hierarchical structure in order to minimize the number of type cast operations whereas the dynamic profiling highlights the hottest dependencies. A major limitation is that HiFPtuner’s configurations are dependent on the tuning inputs, and no accuracy guarantee is provided for untested inputs.

In addition,  the CRAFT tool~\cite{LamHSL13} is a framework that performs an automated search of a program’s instruction space, determining the level of precision necessary in the result of each instruction to pass a user-provided verification routine assuming all other operations are done in high precision such as FP$64$ double precision. However, it can be very time consuming even for very small programs. Let us note that there are other tools oriented to GPU applications ~\cite{AngerdSS17,LagunaWSB19,KotipalliSWLB19} which combine static analysis for casting-aware performance modeling with dynamic analysis for enforcing precision constraints. For an in-depth description and a theoretical description between these tools, we refer the
reader to our survey in~\cite{benkhalifa}.

\section{Conclusion and Future Work}\label{sec6}
In this article, we have  extended our tool POP  relying on a modeling of the propagation
of the errors throughout the code with new optimization criteria in order to obtain a trade-off between, precision, analysis time and memory consumption.  To our knowledge, this is the first work interested in optimizing the results  obtained after the precision tuning phase. The results discussed show that our tool succeeded in limiting the number  of formats,  the number of bits of operations and the number of type conversions between the variables, for the majority of our benchmarks and  with respect to the accuracy requirements given by the user. We shed the light that these results are helpful in the hardware level, especially for some processors that are limited to specific formats.

 In furture work, we aim to adapt our precision tuning tool to generate code in the fixed-point arithmetic. In practice,
 the information provided by POP may be used to generate computations in the fixed-point
 arithmetic with an accuracy guaranty on the results. Also, we are interested in combining POP with other
 tools performing error analysis~\cite{DasKBGT20,DasBGKP20} and code transformation~\cite{DMC15,SaikiFNPT21} tasks in order to better improve the accuracy of the programs. Our technique is also generalizeable to Deep Neural Networks for which it is important to save memory usage and computational resources.
\bibliographystyle{IEEEtran}
\bibliography{IEEEabrv,article}

\end{document}